\documentclass[manuscript]{acmart}
\AtBeginDocument{%
  }

\usepackage{booktabs}
\usepackage{multirow}
\usepackage{array}

\begin{document}
    \begin{teaserfigure}
    \includegraphics[width=\textwidth]{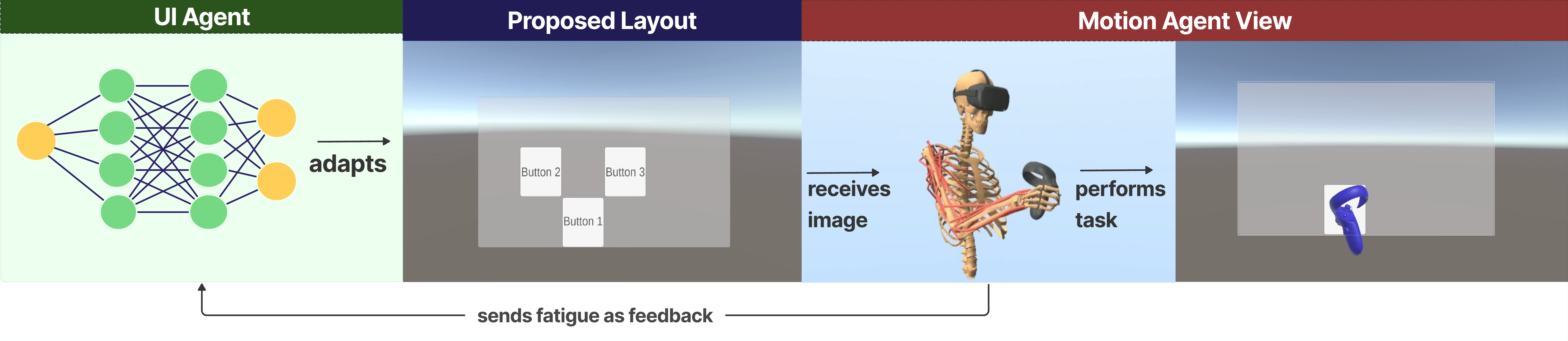}
    \caption{\textbf{Overview of the simulation-driven UI optimization framework.}
    A high-level UI agent proposes discrete interface layouts by selecting positions on a predefined grid.
    Each candidate layout is evaluated by a pre-trained biomechanical motion agent that executes the interaction task in simulation. Muscle-level fatigue is estimated using the 3CC-r model and aggregated to form a reward signal, which enables iterative optimization of UI layouts without labeled data or hand-crafted ergonomic heuristics.}
    \Description{
    Block diagram illustrating a closed-loop simulation-driven UI optimization process.
    A high-level UI agent selects three discrete grid positions to place interface elements.
    These positions are passed to a virtual reality environment, where a pre-trained biomechanical motion agent executes the interaction task.
    The motion agent estimates muscle-level fatigue using a biomechanical model.
    Aggregated fatigue values are returned to the UI agent as feedback, forming a loop that supports iterative refinement of interface layouts.
    }
    \label{fig:teaser}
\end{teaserfigure}

\title {Designing Fatigue-Aware VR Interfaces via Biomechanical Models}

\author{Harshitha Voleti}
\affiliation{%
     \institution{Immersive \& Creative Technologies Lab, Concordia University}
    \country{Canada}}
    \email{voleti.harshitha@gmail.com}

\author{Charalambos Poullis}
\affiliation{%
     \institution{Immersive \& Creative Technologies Lab, Concordia University}
    \country{Canada}}

\renewcommand{\shortauthors}{Voleti et al.}








\begin{abstract}

Prolonged mid-air interaction in virtual reality (VR) can lead to arm fatigue and discomfort, negatively impacting user experience. Incorporating ergonomic considerations into VR user interface (UI) design typically requires extensive human-in-the-loop evaluation. While biomechanical models have been used to simulate human behavior in HCI tasks, their use as surrogate users for ergonomic VR UI design remains underexplored. We propose a framework that leverages biomechanical user models to evaluate and optimize VR interfaces for mid-air interaction. A motion agent is trained to perform button-press tasks in VR under sequential interaction conditions, executing realistic movement strategies and estimating muscle-level effort using a validated three-compartment control with recovery (3CC-r) fatigue model. The simulated fatigue output is then used as a feedback signal for a UI agent that optimizes the spatial layout of UI elements via reinforcement learning (RL), with the goal of minimizing user fatigue. We compare the RL-optimized layout against a manually-designed centered baseline and a Bayesian optimized baseline, and show that fatigue trends predicted by the biomechanical model align with those observed in human users. Furthermore, RL-optimized layout using simulated fatigue feedback resulted in significantly lower perceived fatigue in a subsequent human user study. We further demonstrate the extensibility of our framework through a simulation-based case study involving longer sequential tasks with non-uniform interaction frequencies, examining optimization behavior under increased task complexity. To our knowledge, this is the first work to use simulated biomechanical muscle fatigue as a direct optimization signal for VR UI layout design. Our results highlight the potential of biomechanical user models as effective surrogate tools for ergonomic VR interface design, enabling more efficient early-stage design iteration with reduced reliance on extensive human participation.

\end{abstract}

\maketitle

\section{Introduction}

Designing ergonomic user interfaces for VR remains a persistent challenge. Most VR applications rely on mid-air interaction, often requiring repetitive reaching and sustained arm elevation. These interaction patterns can quickly lead to physical discomfort and muscle fatigue, commonly referred to as the “gorilla arm” effect \cite{laviola20173d}. Such fatigue reduces user comfort, degrades task performance, and limits long-term usability, making physical ergonomics a critical yet frequently under-addressed aspect of VR interface design.

Accounting for physical effort during interface design is inherently difficult. Anticipating how users will move, reach, and sustain postures requires contextual knowledge of task flow and behavior that is typically unavailable during early design stages. Consequently, layouts are often optimized for visual structure or task efficiency, while physical demands are evaluated only after implementation through user studies.

This gap motivates methods that can estimate physical effort during continuous interaction and inform layout decisions prior to human evaluation. Biomechanical user models offer this potential. As muscle-actuated systems, they can simulate realistic motion and incorporate fatigue formulations such as Consumed Endurance (CE) \cite{hincapie2014consumed} and the 3CC-r model \cite{jang2017modeling}. While prior work has used such models to validate motion realism or correlate simulated outcomes with user reports, their use as proactive design tools for optimizing interface layouts remains largely unexplored, particularly in VR, where sustained mid-air interaction dominates.

In this work, we repurpose a biomechanical user model as a surrogate user for ergonomically informed interface design. The simulated user interacts with alternative sequential layouts, generating muscle-level effort signals that are aggregated into cumulative fatigue estimates. These predictions are then used directly to evaluate and optimize interface configurations before human-in-the-loop testing.

Our framework consists of two components: (1) a low-level motion agent that executes interaction tasks and estimates muscle effort, which is aggregated into cumulative fatigue, and (2) a higher-level interface agent that proposes and refines button layouts based on predicted cumulative fatigue. The biomechanical user model is treated as a black-box ergonomic evaluator, enabling systematic exploration of layout alternatives without hand-crafted ergonomic heuristics or auxiliary task-completion rewards. The optimization objective is defined solely in terms of predicted muscle fatigue, penalizing long reaches, awkward postures, and sustained activation patterns.

We evaluate RL-optimized fatigue-aware layouts against a static baseline and a Bayesian optimized baseline in a controlled user study. Results show that simulated fatigue patterns across layouts align with those observed in human participants. Furthermore, RL-optimized layouts reduce perceived physical effort and subjective workload compared to both baselines.

In summary, our contributions are:\\
1. A fatigue-aware VR interface design framework that leverages biomechanical user models as surrogate users to estimate physical effort during continuous interaction at design time. \\
2. A simulation-driven optimization approach that updates interface layouts to minimize predicted cumulative muscle fatigue without relying on heuristic design rules. \\
3. Empirical validation demonstrating that layouts optimized using simulated fatigue reduce perceived physical effort and overall workload in a controlled user study. \\
4. A use-case application of the proposed framework to longer sequential tasks with non-uniform button usage frequencies, examining how cumulative fatigue and usage frequency interact as task complexity increases.

The remainder of this paper is organized as follows. Section~\ref{sec:related_work} reviews related work. Section~\ref{sec:methodology} describes the proposed methodology. Section~\ref{sec:evaluation} presents the evaluation through simulation and a controlled user study. Section~\ref{sec:discussion} discusses the findings, followed by a use case demonstration in Section~\ref{sec:application} and limitations and future work in Section~\ref{sec:limitations_futurework}.

\section{Related Work}
\label{sec:related_work}
\subsection{Evolution of Adaptive User Interfaces}

Designing adaptive user interfaces is a complex optimization problem, and the techniques used to address this challenge have evolved considerably over time. Early adaptive UI systems primarily relied on manually specified rules~\cite{mori2004design}, heuristics~\cite{gobert2019sam}, and explicit user feedback~\cite{hussain2018model}. While these approaches allow designers to encode domain knowledge directly, defining and maintaining explicit rules quickly becomes tedious and difficult to scale.

To reduce manual effort, subsequent work explored optimization-based approaches for interface adaptation, including genetic algorithms~\citep{salem2017user, quiroz2007interactive}, cost-function-driven optimization~\cite{gajos2004supple, sarcar2018ability}, and combinatorial optimization techniques~\citep{cheng2023interactionadapt, lindlbauer2019context, song2025preference}. Although these methods enable systematic exploration of the design space, they often require carefully engineered constraints to ensure feasibility and usability. Defining such constraints is challenging, as these constraints may conflict with one another or overly restrict the solution space.

Data-driven approaches later leveraged observed user behavior to infer preferences and adaptation strategies. Early machine learning methods~\citep{langley1997machine, mezhoudi2013user, pejovic2014interruptme,rathnayake2019framework} enabled automatic adaptation from interaction data, and more recent work has explored deep neural networks for interface optimization~\cite{ soh2017deep,duan2020optimizing}. However, these methods often require large datasets, limiting their practicality in interactive VR settings.

To address data efficiency concerns, online learning approaches such as multi-armed bandits and Bayesian optimization have gained attention. Bandit-based methods iteratively update adaptation strategies by balancing exploration and exploitation based on observed feedback, allowing systems to adapt effectively with limited data~\citep{kangas2022scalable, lomas2016interface}. Similarly, Bayesian optimization treats interface adaptation as a black-box optimization problem, using probabilistic surrogate models to efficiently search for high-performing designs with few evaluations~\citep{dudley2019crowdsourcing, koyama2014crowd}. While both approaches are effective in low-data settings, they typically rely on sparse, per-design evaluations and are sensitive to noise in the feedback signal, which can affect their ability to resolve subtle performance differences across design alternatives.

To address these limitations, RL has been explored as a mechanism for adaptive interface design, as it enables optimization over long-term interaction sequences by maximizing cumulative reward. RL-based AUIs have demonstrated the ability to adapt layouts or interaction strategies over time, accounting for temporal dependencies in user behavior \citep{gebhardt2019learning, todi2021adapting, langerak2024marlui, evangelista2022auit, gaspar2024reinforcement}. While these approaches establish RL as a powerful tool for interface optimization, they primarily focus on task efficiency or user preference and do not explicitly consider physical ergonomics as an optimization objective. Crucially, physical effort and fatigue are not directly observable from interaction traces alone, making them difficult to optimize without an explicit physiological or user effort model.

\subsection{Ergonomic Considerations in XR Interaction}

Physical ergonomics has received increasing attention in XR interface design, but remains underexplored compared to cognitive and contextual factors, particularly as an optimization signal for adaptive layouts. Prolonged mid-air interaction in VR often involves extended reach, repetitive arm movements, and overhead gestures, which can cause discomfort and cumulative fatigue. Early ergonomic approaches aimed to retarget interface elements to more physically comfortable locations using spatial warping and optimization techniques~\cite{montano2017erg}. XRgonomics~\cite{evangelista2021xrgonomics} provided a more systematic framework, estimating ergonomic cost using metrics such as RULA scores, cumulative effort, and muscle activations derived from biomechanical models. However, these approaches primarily focus on static postures or discrete evaluation points, failing to account for fatigue accumulation during extended, continuous interactions. More recent work has begun introducing fatigue-aware interaction strategies. For example,
AlphaPIG \cite{li2025alphapig} continuously adjusts gesture interaction parameters based on a real-time fatigue model so that users receive progressively more assistance only as physical fatigue increases, thereby prolonging comfortable interaction without manual tuning. Such approaches enable designers to explore fatigue-aware interaction strategies systematically, but generally focus on interaction techniques rather than interface layout, leaving an opportunity to optimize UI placement based on physical strain.

\subsection{Fatigue Assessment in Virtual Reality}

Accurate fatigue assessment is central to ergonomics-aware design. In VR, prolonged mid-air interaction can produce significant arm fatigue. Various methods exist to quantify this, including physiological measurements such as blood flow, heart rate, and oxygen consumption, as well as subjective self-report instruments such as the Borg Rating of Perceived Exertion (RPE), NASA-TLX, and Likert-scale questionnaires. Although these methods provide valuable insights into user experience, they typically require continuous monitoring or post-hoc reporting, limiting their suitability for real-time or design-time interface adaptation. Computational models offer an alternative for fatigue estimation. One widely used approach is Consumed Endurance (CE) \cite{hincapie2014consumed}, which estimates fatigue based on Rohmert’s endurance time curve and correlates with subjective exertion ratings. However, CE assumes that interactions below a fixed percentage of maximum strength do not contribute to fatigue and does not explicitly model recovery during rest, which can lead to inaccurate estimates in interactive scenarios. The 3CC fatigue model addresses these limitations by providing a fine-grained, interaction-oriented framework suitable for HCI applications \cite{xia2008theoretical}, and has been shown to predict perceived exertion in mid-air tasks with strong correspondence to Borg CR10 ratings \cite{jang2017modeling}. It was later extended to 3CC-r to better support intermittent interactions by introducing an explicit recovery multiplier, validated against joint-specific perceived fatigue \cite{looft2018modification}. More recently, hybrid models such as NICER \cite{li2024nicer} combine torque-based estimation with empirically measured muscle contraction and recovery terms, improving prediction of above-shoulder fatigue. However, these metrics are primarily evaluated as post-hoc analytical tools rather than as actionable signals for interface design.

\subsection{Biomechanical Modeling in HCI} 

As discussed, machine learning optimization often depends on large amounts of data, which can be difficult or costly to obtain. Biomechanical simulations offer a promising alternative by generating rich, controlled datasets that capture human motion and physical effort. In HCI, such models have gained attention for their ability to objectively estimate muscle-level effort and fatigue, providing a more precise complement to subjective or physiological assessments \cite{bachynskyi2014motion}. Early biomechanical models relied heavily on motion data. This involved mapping physical to virtual markers, scaling the model to match the subject’s dimensions, computing joint angles through inverse kinematics, and estimating muscle activations \cite{bachynskyi2015informing}.In contrast to these motion-driven approaches, Cheema et al.~\cite{cheema2020predicting} demonstrated that torque-driven simulations could predict fatigue and task performance in silico, bypassing the need for human participants. Later work showed that RL could train torque-based models to reproduce human-like movement strategies that follow Fitts’ Law \cite{fitts1954information} and the Two-Thirds Power Law \cite{lacquaniti1983law} \cite{fischer2021reinforcement}. Building on these advances, Ikkala et al.~\cite{ikkala2022breathing} developed muscle-actuated biomechanical user models capable of performing a wide range of HCI tasks, enabling more realistic simulations of motor control and fatigue. These models have also been applied to diverse interaction scenarios \citep{hetzel2021complex, moon2024real, miazga2025increasing}. Most recently, Fischer et al.~\cite{fischer2024sim2vr} introduced Sim2VR, integrating biomechanical simulation with VR applications, which allows simulated users to train and interact in immersive environments.

While prior work has demonstrated that biomechanical simulations can execute and analyze interactive VR tasks, their outputs have primarily been used for validation, behavioral analysis, or performance prediction. The design implications of simulated muscle activity have remained largely unexplored. In this work, we shift the role of biomechanical modeling from analysis to design optimization. Specifically, we incorporate muscle-level fatigue estimates as the objective signal in an RL pipeline that iteratively refines UI placement. Rather than evaluating completed interfaces post hoc, the simulation directly informs layout generation, enabling ergonomic considerations to shape interface structure during the design process without requiring additional human data collection.
\section{Methodology}
\label{sec:methodology}

We formulate VR interface layout optimization as a hierarchical simulation problem. A low-level motion agent, implemented as a muscle-actuated biomechanical user model, executes interaction tasks and produces muscle activation signals. These signals are aggregated into fatigue estimates, which define the objective optimized by a higher-level UI agent that updates interface layouts.

We first present the overall simulation–optimization framework and its integration with the SIM2VR pipeline. We then describe the interaction task used to evaluate candidate layouts, followed by details of the motion agent responsible for executing the task and estimating fatigue. Finally, we formalize the UI agent and its reinforcement learning formulation for interface layout optimization.

\subsection{Framework}

\begin{figure}[htbp] 
    \centering
    \includegraphics[width=0.8\textwidth]{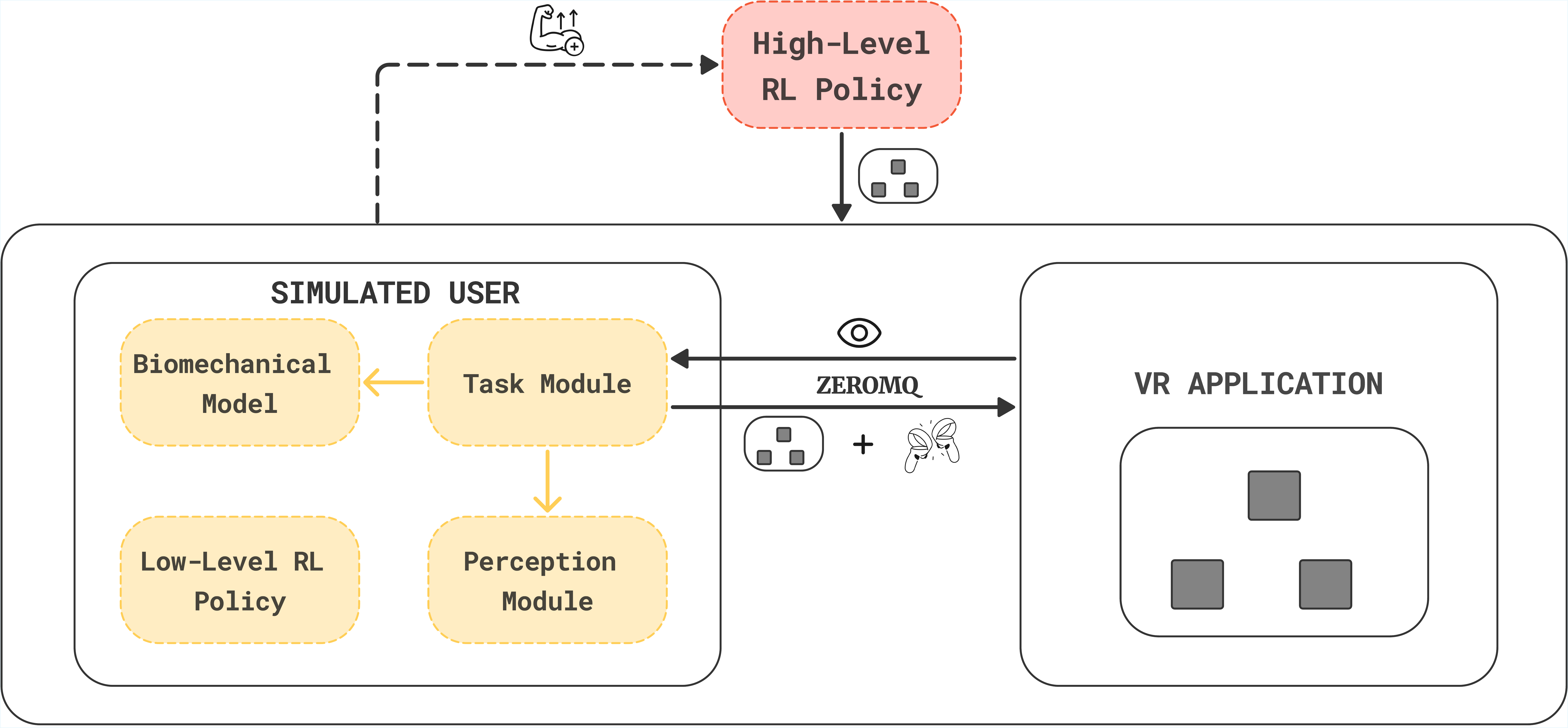} 
\caption{\textbf{Overview of the proposed hierarchical framework}.
A high-level UI agent proposes discrete interface layouts and evaluates them using cumulative fatigue feedback from a simulated user. 
Each candidate layout is instantiated in the Unity-based VR application through the SIM2VR task module. Rendered observations are provided to the simulated user's perception module, which is fed into a learned control policy that outputs muscle activation signals. 
A low-level motion agent controls a biomechanical model to execute the interaction sequence under the proposed layout. Muscle activations are logged during execution and converted into fatigue estimates. 
The cumulative fatigue over the completed sequence is returned to the UI agent as a reward signal, together with the resulting layout state, to guide subsequent layout optimization.}

    \Description{
    Block diagram illustrating a hierarchical VR interface optimization pipeline.
    A high-level UI agent proposes candidate button layouts.
    Each layout is evaluated by a simulated user implemented through the SIM2VR pipeline.
    Within this pipeline, a low-level motion agent receives rendered observations from a Unity-based VR application, processes them through a perception module and learned control policy, and outputs muscle control signals applied to a biomechanical arm model.
    During task execution, muscle activations are logged and converted into fatigue estimates.
    Aggregated fatigue values are returned to the UI agent, forming a closed feedback loop for iterative layout optimization.
    }

    \label{fig:framework}
\end{figure}

Our framework builds on SIM2VR~\cite{fischer2024sim2vr}, which integrates a muscle-actuated biomechanical user model with Unity-based VR applications. In SIM2VR, the model is trained via RL in Python and communicates with the VR environment through a ZMQ interface, enabling goal-directed arm movements and interaction with virtual objects. We extend this pipeline by introducing a UI agent that proposes candidate interface layouts (Fig.~\ref{fig:framework}). For each layout, the configuration is transmitted to the VR environment using the existing SIM2VR modules. The biomechanical motion agent then executes the task on the updated interface. During execution, muscle activations are recorded at each timestep and used to compute cumulative fatigue estimates. These estimates are returned to the UI agent as the optimization signal for subsequent layout updates.

\subsection{Task Description}
\label{sec:task-description}
We define a sequential button-selection task to study repeated reaching and selection movements. A $0.64\,\mathrm{m} \times 0.36\,\mathrm{m}$ UI canvas is positioned $0.58\,\mathrm{m}$ in front of the user’s head, approximating a comfortable mid-air reach for seated or standing users. The dimensions also account for the biomechanical model’s lack of neck and head movement, ensuring that the UI remains within the field of view while staying within reach. Three square buttons are displayed sequentially, with the next button appearing only after the previous one has been pressed. Buttons were modeled as square targets of size $0.10\,\mathrm{m} \times 0.10\,\mathrm{m}$, consistent with prior VR fatigue and reaching studies that employ targets of comparable physical extent to elicit sustained mid-air interaction while avoiding excessive precision demands (e.g., \cite{jang2017modeling}). This size reflects a common design choice in fatigue-oriented interaction tasks rather than an optimized UI parameter. Users complete the task by selecting all buttons with a VR controller. Buttons are axis-aligned, meaning their edges are parallel to the canvas axes, simplifying reach computation and ensuring consistent placement. This minimal sequential selection task isolates fatigue arising primarily from arm movements, minimizing confounding factors such as visual search or decision-making strategies.

\subsection{Motion Agent}
\label{sec:motion-agent}
The motion agent models human behavior during the button-selection task using the upper-body, right-handed biomechanical model \textit{MoblArmsWrist} \cite{saul2015benchmarking}, as implemented in \cite{fischer2024sim2vr}. We focus on this model because our primary interest is shoulder fatigue.
\textit{MoblArmsWrist} provides a physiologically grounded representation of arm motion, explicitly modeling muscle activations. It includes shoulder, elbow, and wrist joints, and the muscles actuating these joints, resulting in a total of 7 DOFs and 32 muscles.

\subsubsection{Overview}

\begin{figure}[htbp] 
    \centering
    \includegraphics[width=0.8\textwidth]{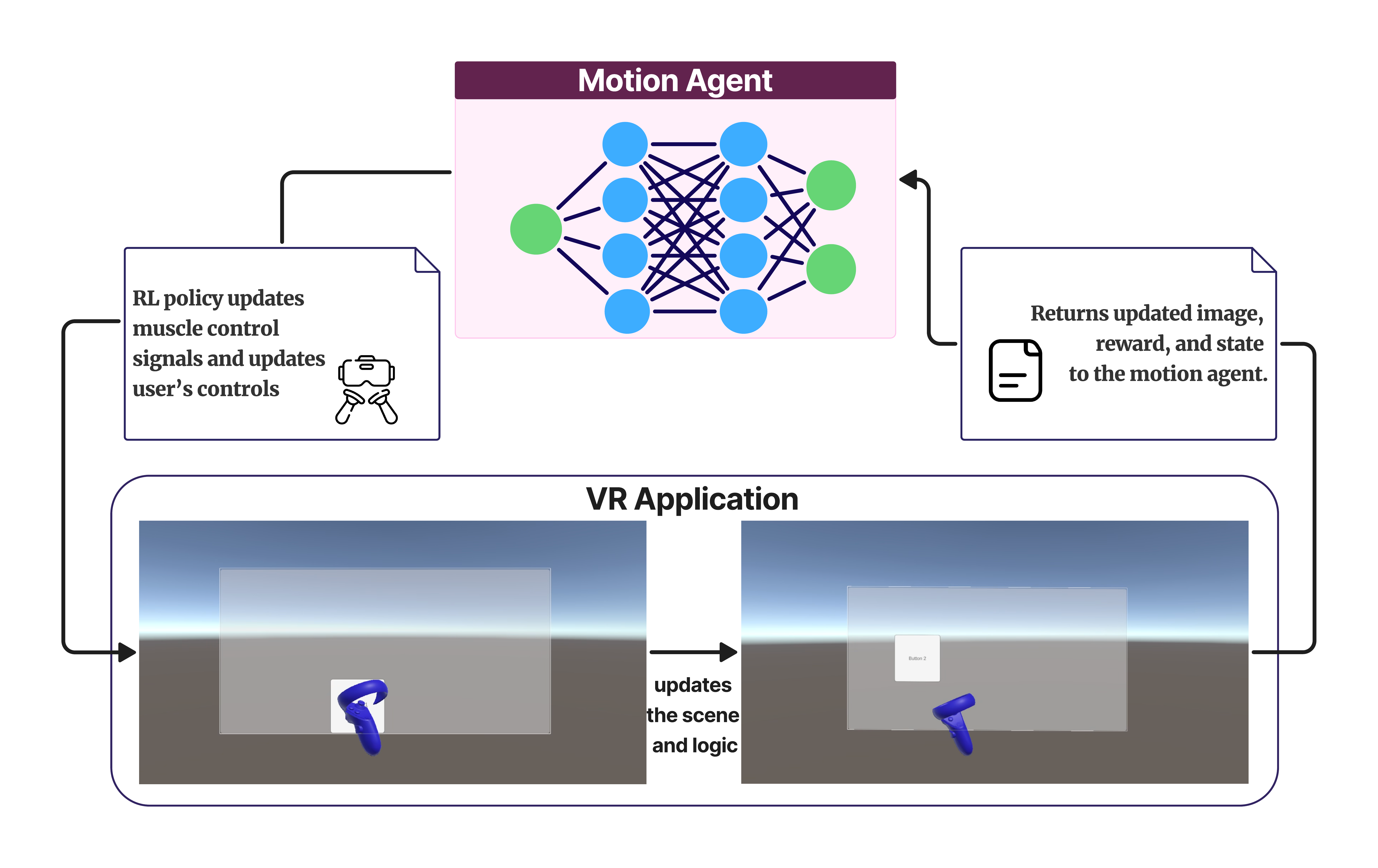} 
    \caption{\textbf{Overview of the motion agent used to simulate interaction.}
    The motion agent generates muscle control signals using an RL policy and interacts with the VR application, which updates the scene and returns rendered observations, reward, and state information. Muscle activations produced during task execution are logged and later converted into fatigue estimates, which are used to evaluate interface layouts.}
    \Description{
    Diagram illustrating the motion agent control loop used to simulate human interaction.
    The motion agent receives RGB-D images and proprioceptive state information from a Unity-based VR application.
    These observations are processed by a learned policy that outputs continuous muscle control signals applied to a biomechanical arm model.
    As the arm moves and presses UI elements, muscle activations are logged and later converted into fatigue estimates.
    The environment updates the scene and returns new observations, allowing the loop to repeat over the interaction sequence.
    }
    \label{fig:motion_agent}
\end{figure}

The implementation of the motion agent follows the same underlying structure as in \cite{fischer2024sim2vr}. The agent receives an image input of the VR application as a 120×80 pixel RGB-D array along with proprioceptive state features, which are jointly encoded using a convolutional neural network (CNN). During training, the agent is guided by a reward objective that encourages successful button selection while promoting efficient movement toward the target. The agent outputs continuous muscle control signals in the range $[-1, 1]$, which are used to actuate the biomechanical arm model. (Fig.~\ref{fig:motion_agent})

Similar to SIM2VR, the motion agent is not scaled to represent a specific participant; instead, it functions as a generic, physically plausible surrogate for upper-limb motion rather than a personalized user model.

\subsubsection{Reward Design}
\label{sec:motionagent_rewardstruct}
Prior work by \cite{selder2025demystifying} provides guidance on reward design for biomechanical simulations. Accordingly, the reward function is structured to include three components: a task-specific reward, a distance-based term, and an effort regularization term.

\begin{itemize}
\item \textbf{Task-specific reward:}  
A sparse reward is provided only when the task is successfully completed, i.e., when a button is pressed. Both the controller and the button are equipped with colliders, and a button press is registered when the controller collider intersects the button collider within a predefined proximity. The button collider is placed at the center of the button, while the controller collider is placed on the trigger of the controller, ensuring that successful interaction requires alignment between the controller and button centers.

During motion agent training, episodes terminate after a fixed time limit of 60 seconds, and buttons are spawned at random positions on the UI canvas. The agent is tasked with pressing as many buttons as possible within the allotted time, encouraging generalizable reaching behavior across the full interface space. 

Each successful button press yields a reward of 5:

\[
r_{\text{task}} =
\begin{cases}
5, & \text{if the button is pressed},\\[2pt]
0, & \text{otherwise}.
\end{cases}
\]

\item \textbf{Distance term:}
A dense reward is computed at each timestep based on the Euclidean distance between the controller and the target ($d$). Given the fixed 0.58 m distance to the UI canvas, distances remain small, thus an exponential formulation is used to provide stronger gradients near the target and a smoother shaping signal during reaching:
\[
r_{\text{d}} = e^{-\mathrm{d}} - 1.
\]

Fig.~\ref{fig:distance_reward} illustrates the shape of the distance-based shaping term used during training.

\begin{figure}[t]
  \centering
  \includegraphics[width=0.5\linewidth]{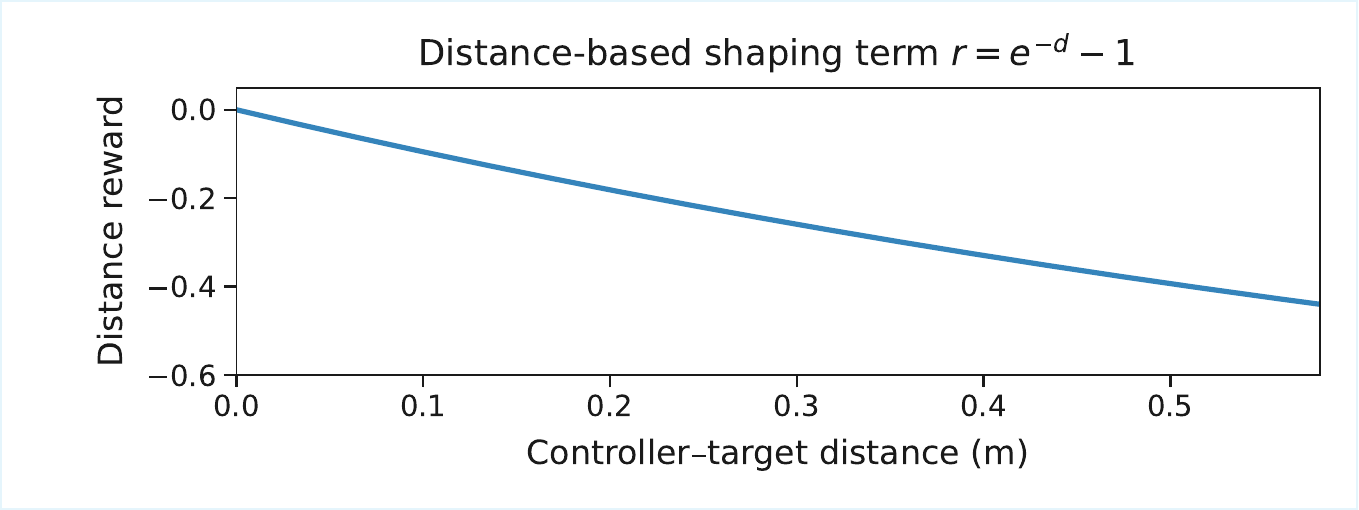}
  \caption{Distance-based shaping term used for reward design.}
  \label{fig:distance_reward}
\end{figure}

\item \textbf{Effort regularization:}  
The effort term \(\delta t\) is subtracted directly from the total reward to penalize higher physical effort, thereby discouraging unnecessary or inefficient movements \cite{ikkala2022breathing,fischer2024sim2vr}. It represents the instantaneous effort cost \(C_{\mathrm{eff}}\) computed by the biomechanical model and is multiplied by a constant to scale its impact:
    \[
    \delta t = 0.01 \cdot C_{\mathrm{eff}}
    \]
    where \(C_{\mathrm{eff}}\) is the instantaneous effort cost produced by the biomechanical fatigue model.
\end{itemize}

The total reward of the motion agent for each timestep is:

\[
r = r_{\text{task}} + r_{\text{d}} - \delta t
\]

\subsubsection{Training Setup}

We train the motion agent using Proximal Policy Optimization (PPO), following the training configuration reported in SIM2VR~\cite{fischer2024sim2vr} to ensure consistency with prior work.
The policy is a multi-input actor–critic network with two fully connected layers of 256 units each and LeakyReLU activations. Continuous muscle activations are output with Tanh scaling, and the feature extractor processes both RGB-D visual input and state variables.

\subsubsection{Fatigue Calculation}

SIM2VR provides several models for estimating effort and fatigue in biomechanical simulations, including CE~\cite{hincapie2014consumed} and variants of the 3CC model. In this work, we use the Three Compartment Controller with recovery (3CC-r) model, which explicitly accounts for both fatigue accumulation and recovery over time. The 3CC-r model maintains internal muscle states corresponding to active, resting, and fatigued capacity, and computes a scalar \(C_{\mathrm{eff}}\) based on the difference between target muscle activations and currently available capacity. In our framework, \(C_{\mathrm{eff}}\) serves as the primary fatigue signal for UI optimization, capturing the physical demand experienced during interaction. Although the model also outputs a muscle fatigue index \(M_F\), prior work has shown that \(M_F\) can saturate under prolonged or near-static loading~\cite{cheema2020predicting}. Because mid-air VR interaction often involves sustained reaches and postures, we rely on \(C_{\mathrm{eff}}\) as a more sensitive measure of fatigue for layout comparison. The choice of \(C_{\mathrm{eff}}\) as the fatigue signal allows us to compare interface layouts based on predicted user fatigue without requiring absolute calibration to human fatigue levels. Its role in guiding layout optimization is described in Section~\ref{sec:ui-agent}.

\subsubsection{Motion Agent Validation}
The trained motion agent reliably completes the button-selection task across most of the UI canvas, making it suitable for use as a fatigue evaluator during layout optimization. Failures primarily occur when buttons are placed at the extreme left edge of the canvas, which is consistent with the reach limitations of the right-handed biomechanical model used in this work.

Since the UI agent relies on relative fatigue comparisons between candidate layouts rather than absolute task success rates, these limitations define the effective design space explored during optimization. Layouts that require repeated interaction in regions beyond the model’s reachable workspace naturally incur higher predicted fatigue or task failure and are therefore disfavored by the UI agent.

\subsection{UI Agent}
\label{sec:ui-agent}

\subsubsection{Role and Problem Formulation}

\begin{figure}[htbp] 
    % \centering
    \includegraphics[width=0.75\textwidth]{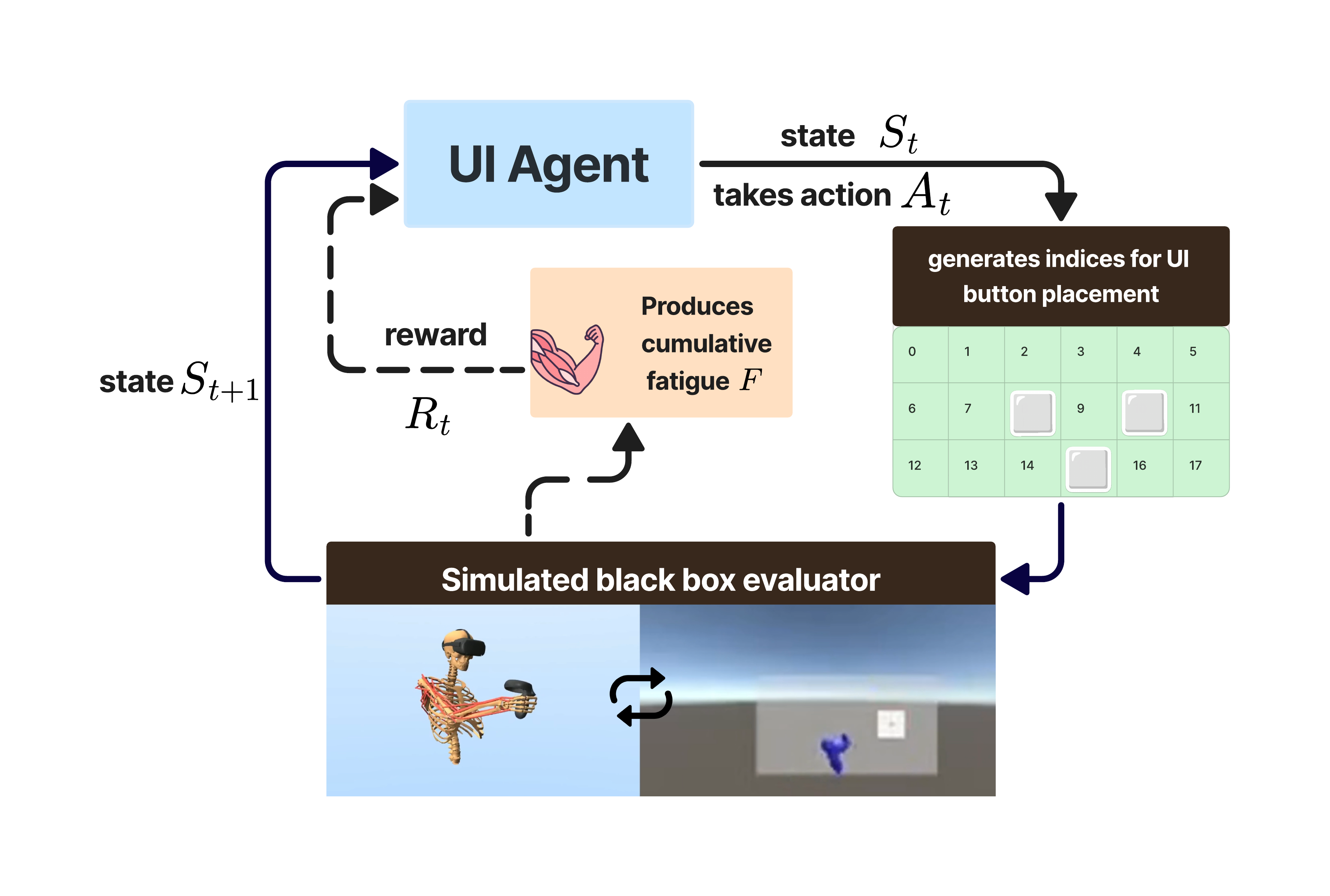} 
    \caption{\textbf{Overview of the UI agent responsible for interface layout optimization.}
    The agent proposes discrete grid-based button layouts, which are instantiated in the VR environment and evaluated by the motion agent. Aggregated fatigue is returned as a reward signal, along with the next state of the UI agent defined by the grid coordinates of all button positions.}
    \Description{
    Diagram showing the UI agent workflow.
    The UI agent selects grid-based positions for multiple UI buttons and instantiates the resulting layout in the virtual environment. A simulated user then executes the interaction sequence after which the UI agent receives aggregated fatigue feedback along with the next state, defined by the x,y coordinates of the current button positions. This feedback is used to evaluate and refine future layout proposals.
    }
    \label{fig:ui_agent}
\end{figure}

The UI agent operates at the interface design level and is responsible for proposing button layouts that minimize predicted user fatigue during interaction. Unlike the motion agent, which interacts with the VR environment through embodied movement and physics-based control, the UI agent interacts with the VR system by parametrically modifying the interface layout. Specifically, it proposes button placements that are instantiated in the Unity-based VR application, after which the simulated user perceives and interacts with the updated layout. The UI agent itself does not execute physical actions, but treats the simulated user as a black-box evaluator that returns fatigue estimates for a given layout. The overall UI agent workflow is illustrated in Fig.~\ref{fig:ui_agent}.

We formulate UI layout optimization as a discrete episodic RL problem. Each episode corresponds to evaluating a single UI layout for the fixed sequential button-selection task described in Section~\ref{sec:task-description}. The episode terminates once the simulated user has attempted the full button sequence, and the cumulative fatigue serves as the sole performance signal. This formulation enables automated exploration of interface design alternatives based purely on predicted physical comfort, without relying on heuristic ergonomic rules.

Although the layout space is discrete and low-dimensional, we adopt RL to handle stochastic, simulation-based evaluation in which fatigue estimates vary due to biomechanical dynamics and recovery effects. This formulation allows the UI agent to learn under noisy ergonomic feedback while incorporating soft constraints via reward penalties for infeasible layouts (e.g., unreachable or overlapping button placements), rather than explicitly restricting the action space.

\subsubsection{Action Space}
The UI canvas has dimensions $0.64\,\mathrm{m} \times 0.36\,\mathrm{m}$ and contains three square buttons of size $0.10\,\mathrm{m} \times 0.10\,\mathrm{m}$. To keep the optimization problem tractable while ensuring physically valid and visually distinct placements, and to reflect realistic UI layouts commonly used in VR applications, we discretize the canvas into a $3 \times 6$ grid, yielding 18 non-overlapping cells. Each cell corresponds to a valid button placement that lies fully within the canvas boundaries.

We adopt a discrete action space for two primary reasons. First, interface layout optimization is fundamentally a combinatorial design problem rather than a continuous control task. Each action represents the selection of grid-based positions for multiple UI elements, and evaluating any proposed layout requires executing a full biomechanical simulation, making reward signals both computationally expensive and stochastic. By discretizing the canvas into semantically meaningful grid cells, we restrict the search to structurally distinct configurations, thereby improving sample efficiency under costly rollouts. Second, discretization avoids degenerate solutions common in continuous optimization, such as near-overlapping buttons or infinitesimal spatial adjustments that produce negligible ergonomic differences but introduce instability in policy learning.

The UI agent’s action is a discrete vector
\[
\mathbf{a} = (a_1, a_2, a_3),
\]
where each $a_i \in \{0, \dots, 17\}$ specifies the grid cell index for button $i$. All three button positions are proposed simultaneously at the beginning of an episode. Layouts that assign multiple buttons to the same grid cell are penalized through the reward function, discouraging overlapping or ambiguous configurations without imposing hard constraints on the action space.

\subsubsection{Observation Space}

The observation provided to the UI agent consists of the current layout configuration expressed as the $(x,y)$ grid coordinates of each button. These coordinates are normalized to $[0,1]$ based on the canvas dimensions. The UI agent does not receive any direct information about the user’s kinematics, task state, or workspace geometry. This deliberately minimal observation design reflects realistic interface design settings, where designers must reason about ergonomic quality without access to detailed kinematic measurements of the user, and instead rely on aggregate comfort-related feedback.

\subsubsection{Reward Function}

The objective of the UI agent is to minimize cumulative physical effort experienced by the simulated user over the entire interaction sequence. During task execution, the motion agent produces a scalar instantaneous effort cost $C_{\mathrm{eff}}(t)$ at each timestep using the 3CC-r fatigue model (Section~\ref{sec:motion-agent}). These values are accumulated over time to compute sequence-level fatigue. For the three-button sequential task, we define the cumulative effort associated with selecting button $i$ as

\[
F_i = \sum_{t \in \mathcal{T}_i} C_{\mathrm{eff}}(t),
\]

where $\mathcal{T}_i$ denotes the set of timesteps corresponding to the reach or transition to button $i$ (from the resting posture for $i=1$ and from the previously selected button for $i>1$).

The episode-level reward is then defined as

\[
r = - \sum_{i=1}^{3} F_i .
\]

Each episode begins from a fixed initial posture, and the motion agent’s internal fatigue state is reset between episodes so that rewards reflect fatigue accumulated within a single interaction sequence. This formulation directly aligns the optimization objective with ergonomic comfort: layouts that require longer reaches, awkward postures, or sustained muscle activation receive lower rewards. No additional task-completion or heuristics are included in the reward, allowing the UI agent to discover fatigue-efficient layouts autonomously.

To ensure robustness, we discourage infeasible layouts through reward penalties rather than explicitly restricting the action space. Instead of masking invalid actions, we allow the UI agent to explore the full layout space and assign large penalties to unreachable or ill-formed configurations. If the motion agent fails to press a button within the first 15 seconds of an episode, a large fatigue penalty is assigned and the episode terminates early. Similarly, layouts that assign multiple buttons to the same grid cell incur a penalty and trigger early termination before deployment in VR. This reward-based constraint formulation encourages the UI agent to autonomously learn reachable, well-formed layouts while preserving fatigue minimization as the primary optimization objective.

\subsubsection{Training Setup}

We train the UI agent using PPO. The policy network is a lightweight fully connected actor–critic model with two hidden layers of 128 units each and LeakyReLU activations. A custom feature extractor first normalizes the button coordinate inputs using layer normalization and then projects them into a 64-dimensional latent representation via a linear layer with LeakyReLU activation, before policy and value prediction. Training is performed until policy convergence using multiple parallel environments. 

This setup enables the UI agent to iteratively refine layout proposals based on predicted physical effort, providing a principled mechanism for exploring fatigue-aware interface designs. Because layout proposals and fatigue estimates are evaluated at the level of individual button placements, the learned policies can be post hoc analyzed to identify spatial patterns associated with lower predicted physical demand, offering opportunities for design insight beyond black-box optimization.

\section{Evaluation}
\label{sec:evaluation}

Two baseline UI configurations were considered for comparison. The first is a static, hand-designed layout with all buttons positioned at the center of the UI canvas, reflecting a conventional and intuitive design. The second is a Bayesian-optimized layout, generated using Bayesian Optimization (BO) to minimize predicted user fatigue. BO treats the mapping from candidate UI layouts to predicted fatigue as a black-box objective and is commonly used for data-efficient interface adaptation when evaluations are expensive or noisy, providing a natural optimization-based baseline for this study. (Section~\ref{sec:related_work}).

BO is implemented using Ax \cite{olson2025ax}. The optimization is initialized with 15 Sobol-sampled layouts, followed by 250 BO iterations. Each candidate layout is evaluated by the motion agent, which executes the sequential button-selection task and returns the sequence-level accumulated effort
\[
F = \sum_{t=1}^{T} C_{\mathrm{eff}}(t),
\]
i.e., the sum of the instantaneous effort cost $C_{\mathrm{eff}}(t)$ produced by the 3CC-r model over all timesteps in the sequence. This value is used as feedback to the optimizer. This setup enables systematic comparison between conventional central placement and layouts discovered through simulation-driven fatigue-aware optimization.

\subsection{Simulation Results}

\begin{figure}[t]
    \centering
    \begin{minipage}[b]{0.32\linewidth}
        \centering
        \includegraphics[width=\linewidth]{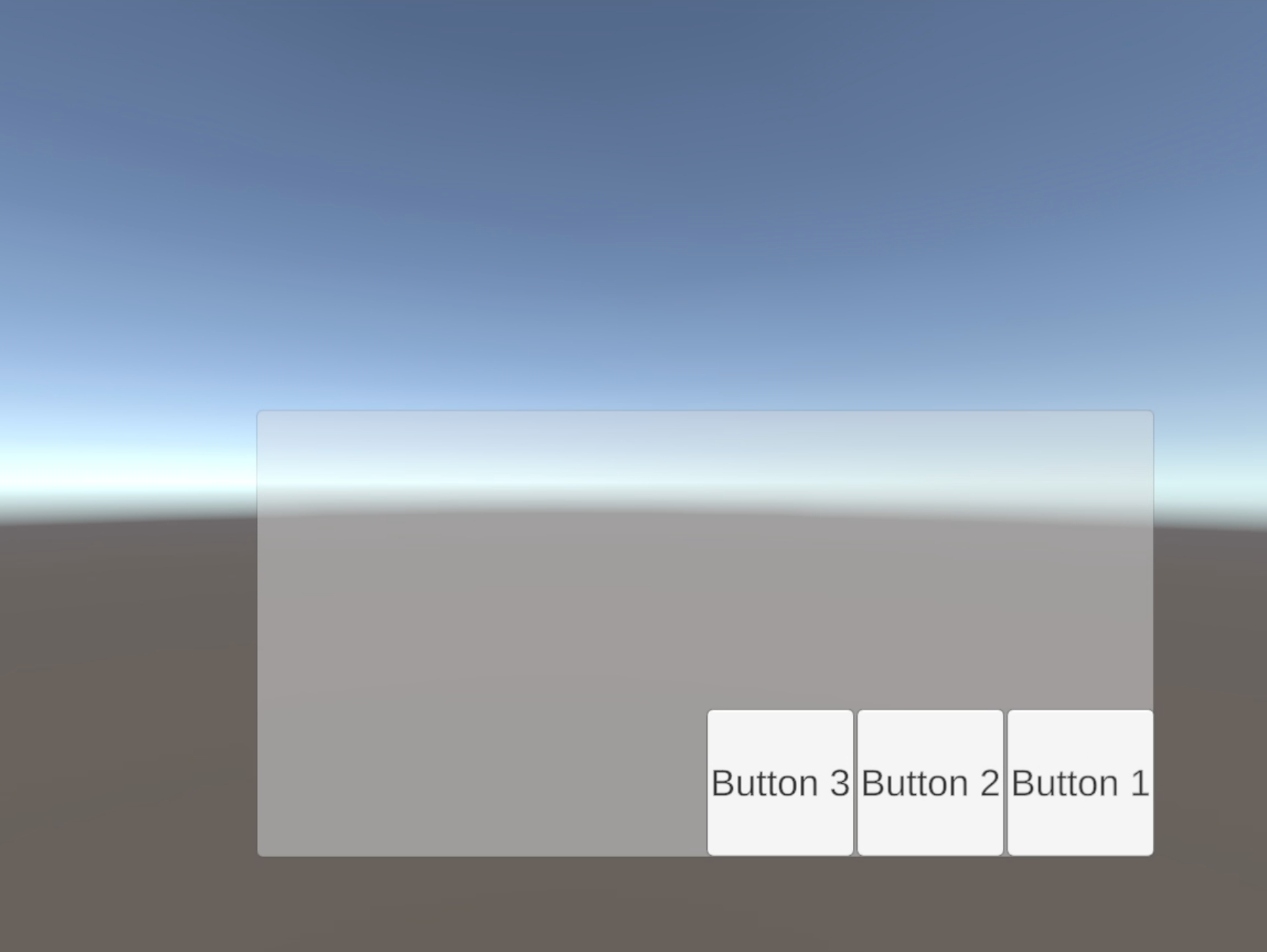}\\
        \small (a) RL
    \end{minipage}\hfill
    \begin{minipage}[b]{0.32\linewidth}
        \centering
        \includegraphics[width=\linewidth]{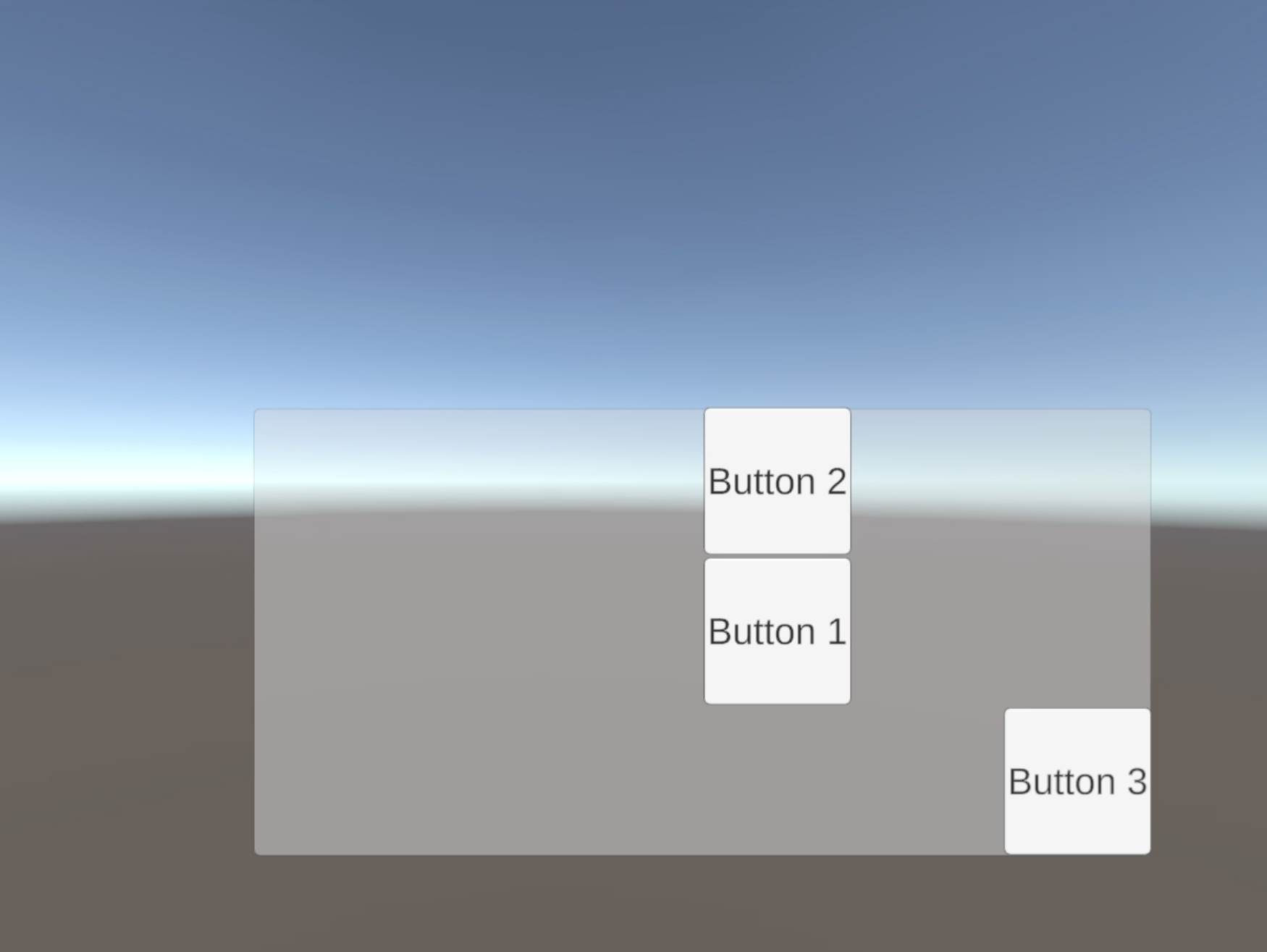}\\
        \small (b) BO
    \end{minipage}\hfill
    \begin{minipage}[b]{0.32\linewidth}
        \centering
        \includegraphics[width=\linewidth]{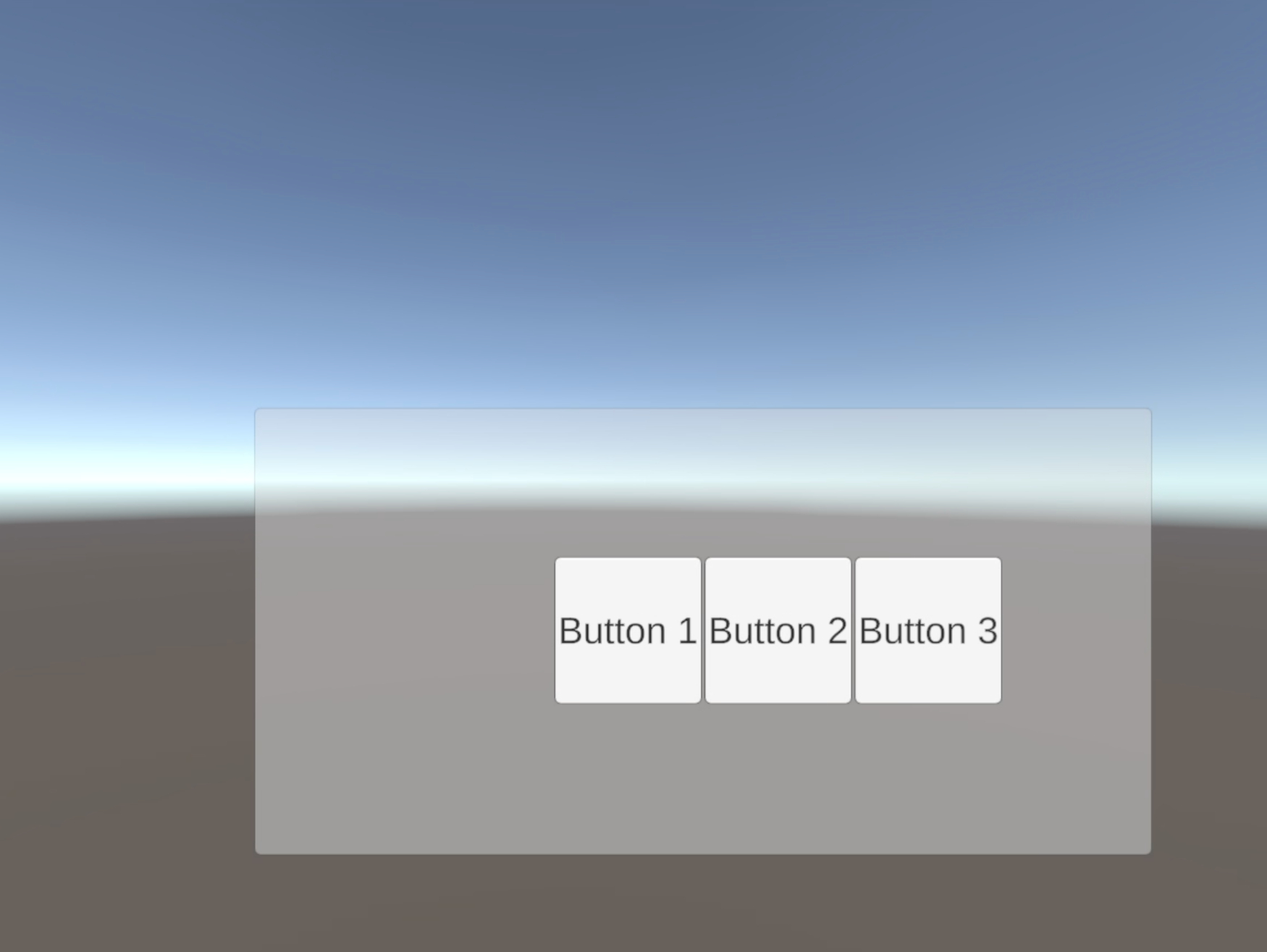}\\
        \small (c) Static
    \end{minipage}
    \caption{\textbf{UI configurations evaluated in the study.}
    (a) RL-based layout optimized using biomechanical fatigue feedback,
    (b) Bayesian optimization (BO) layout minimizing accumulated effort,
    and (c) static layout with centrally placed buttons.}

    \Description{
    Three UI layout visualizations are shown.
    The left image shows the RL-based layout, with three buttons placed close together near the right edge of the grid.
    The middle image shows the Bayesian-optimized layout, where buttons are spread across different grid cells.
    The right image shows the static baseline layout, with three buttons placed centrally on the grid.
    }
    \label{fig:ui_layouts}
\end{figure}

We compare layouts produced by the RL agent against the static baseline and the BO-based baseline. All layouts are defined over a discrete $3 \times 6$ grid on a $0.64\,\mathrm{m}\times 0.36\,\mathrm{m}$ UI canvas. The resulting RL, BO, and static layouts are shown in Fig.~\ref{fig:ui_layouts}.

\paragraph{RL Agent} After training, the RL agent consistently selects grid cells 17, 16, and 15 for the sequential button task. These positions correspond to the rightmost section of the grid, which the biomechanical simulation predicts to yield lower sequence-level accumulated effort ($F$) over the interaction sequence. The selected layouts cluster buttons within a contiguous, reachable region of the workspace, resulting in lower $F$ across sequential button transitions.

\paragraph{BO} After performing optimization, we selected the layout recommended by Ax as the best-performing configuration, based on the observed objective values across evaluations. This resulted in a layout placing buttons at grid cells 9, 3, and 17, with a mean accumulated effort of $25.26$ (variance $21.09$), which we used as the BO baseline in the user study.
%can i remove which corresponds to the minimizer of the surrogate model’s posterior mean and accounts for observation noise since i dont know if this is true or not? i dont see it in documentation unless u can provide proof?

% probably needs more information over here.
\paragraph{Static Layout} The static baseline places buttons at grid cells 8, 9, and 10, forming a centrally aligned horizontal configuration on the UI canvas. This layout reflects a common default design choice in VR interfaces, where elements are positioned near the center of the user’s field of view. Because it does not account for user biomechanics or the cost of transitions between successive button selections, it serves as a naive reference for evaluating fatigue-aware layout optimization.

%---------------------------------------------------------------------------
%-------------------------------------------------------------------
\subsection{User Study}

\subsubsection{Setup}
\paragraph{Participants}
Eighteen participants
% \footnote{Sample size was determined via an a priori G*Power analysis for a within-subjects ANOVA (three conditions), assuming $\alpha = .05$, power = .95, and a medium-to-large effect size ($f = 0.40$), yielding a required sample size of 18.}
were recruited to interact with the three UI configurations. All participants were students enrolled in institutions of higher education. The sample included 11 male and 7 female participants, with a mean age of 26.5 years (SD = 2.25). All participants were right-handed, as the simulated user employed in our pipeline is based on a right-handed biomechanical model. Twelve participants reported prior experience with VR, while six had little or no prior exposure.

\paragraph{Apparatus} We used a Meta Quest 2 HMD, powered by a computer running on an Intel Core 11th Gen CPU and equipped with an NVIDIA GeForce RTX 3080 Ti. 

\paragraph{Procedure} Each UI configuration was treated as a separate condition and consisted of 30 sequential button-interaction tasks. Participants interacted with each condition three times, resulting in a total of 90 interaction sequences per condition per participant. The order of UI conditions was counterbalanced using a Latin square design in a within-subject experiment. Participants were given a short break of 30 seconds between interaction sequences and a longer break of one minute after completing each condition, with additional rest available upon request.

During the study, participants were seated on a chair in an office environment and were instructed to maintain an upright posture while minimizing torso movement, in order to match the fixed-torso assumption of the simulated user model. After completing each condition, participants provided subjective feedback by reporting their perceived effort using the Borg CR10 scale as well as workload ratings using the NASA Task Load Index (NASA-TLX). On average, participants required between 45 and 50 minutes to complete the study.

\subsection{Results}

We evaluated the three UI configurations using both simulated effort estimates and subjective fatigue reported by human participants. Data were analyzed using SPSS 31. Distributions were considered approximately normal when Skewness (S) and Kurtosis (K) were within $\pm 1$ \citep{hair2019multivariate, mallery2000spss}; otherwise, a log transformation was applied. 

\subsubsection{Predicted Effort and Perceived Exertion}

\paragraph{Simulated User}
We analyzed the accumulated effort predicted by the biomechanical model across the three UI layouts. Since the UI agent is optimized at the sequence level, we report sequence-level accumulated effort ($F$), defined as the sum of the instantaneous effort cost $C_{\mathrm{eff}}(t)$ over all timesteps in a sequence. Each configuration was evaluated over 30 trials using identical sequences, and mean $F$ values were reported. In the simulation, muscle states were reset between sequences to isolate the effort required for a single sequence. In contrast, participants completed sequences consecutively and reported perceived exertion after each condition. Although these measures capture different timescales, layouts with lower predicted $F$ per sequence are expected to result in lower perceived exertion when repeated. A Friedman test revealed a significant effect of UI configuration on predicted accumulated effort ($\chi^2(2) = 48.26$, $p < 0.001$). Post-hoc comparisons indicated that the RL-based UI yielded significantly lower $F$ than both the Static and BO-based UIs, while the Static UI yielded lower $F$ than the BO-based UI.

\paragraph{User Study}
For the human participants (N = 18), we applied the same non-parametric analysis to perceived exertion measured using the Borg CR10 scale. A Friedman test showed a significant effect of UI configuration on perceived exertion ($\chi^2(2) = 22.90$, $p < 0.001$). Post-hoc comparisons revealed the same ordering across UI conditions: the RL-based UI produced the lowest fatigue, followed by the Static UI, with the BO-based UI yielding the highest fatigue.

Descriptive statistics and post-hoc pairwise Wilcoxon signed-rank comparisons are summarized in Table~\ref{tab:fatigue_wilcoxon}.

Overall, the relative ordering of UI configurations was consistent across simulated effort estimates and subjective fatigue reports, suggesting alignment between the biomechanical model’s predictions and human-perceived exertion.

%------------------- TABLE CREATION ----------------------------------
\begin{table}[t]
\centering

\setlength{\tabcolsep}{6pt}
\renewcommand{\arraystretch}{1.15}

\begin{tabular}{l l l c c l  c c}
\toprule
\textbf{User} &
\textbf{Fatigue Type} &
\textbf{UI} &
\textbf{Mean} &
\textbf{Std.} &
\textbf{Comparison} &
\multicolumn{2}{c}{\textbf{Wilcoxon Signed Rank}} \\
\cmidrule(lr){7-8}
& & & & & & \textbf{Z} & \textbf{p} \\
\midrule

\multirow{3}{*}{Simulated User} &
\multirow{3}{*}{Accumulated effort ($F$)} &
UI\_Static & 30.4513 & 4.0708 & $UI_{Static} < UI_{BO}$ & -2.910 & $ < 0.01$  \\
& & UI\_BO & 33.0482 & 1.3801 & $UI_{RL} < UI_{BO}$ & -4.782 & $ < 0.001$ \\
& & UI\_RL & 22.8955 & 0.7652 & $UI_{RL} < UI_{Static}$ & -4.782 & $ < 0.001$ \\
\addlinespace[2pt]
\midrule

\multirow{3}{*}{User Study} &
\multirow{3}{*}{Borg CR10} &
UI\_Static & 3.639 & 2.2016 & $UI_{Static} < UI_{BO}$ & -2.875 & $ < 0.01$ \\
& & UI\_BO & 5.94 & 1.662 & $UI_{RL} < UI_{BO}$ & -3.598 & $ < 0.001$ \\
& & UI\_RL & 1.889 & 1.6230 & $UI_{RL} < UI_{Static}$ & -3.210 & $ = 0.001$ \\

\bottomrule
\end{tabular}
\caption{Descriptive statistics and Wilcoxon signed-rank post-hoc comparisons of fatigue across UI conditions.}
\label{tab:fatigue_wilcoxon}
\end{table}
%------------------------------------------------------------------------------
\subsubsection{Task Completion Time}

We measure task completion time to examine whether UI layout influences user performance. Since the next button appears only after the current button is successfully pressed, accuracy and error rates are not analyzed separately; errors are implicitly reflected as increased completion time.

A repeated-measures ANOVA was performed on log-transformed task completion time (S = 0.54, K = 0.31). Results revealed a significant effect of UI condition on completion time ($F(2, 34) = 5.92$, $p = 0.006$, $\eta_p^2 = 0.26$). 
Post-hoc comparisons with Bonferroni correction indicated that UI\_BO resulted in significantly longer completion times than both UI\_Static ($p < 0.05$) and UI\_RL ($p < 0.05$), while no significant difference was observed between UI\_Static and UI\_RL.

\subsubsection{NASA-TLX}
\begin{figure}[t]
    \centering
    \includegraphics[width=\linewidth]{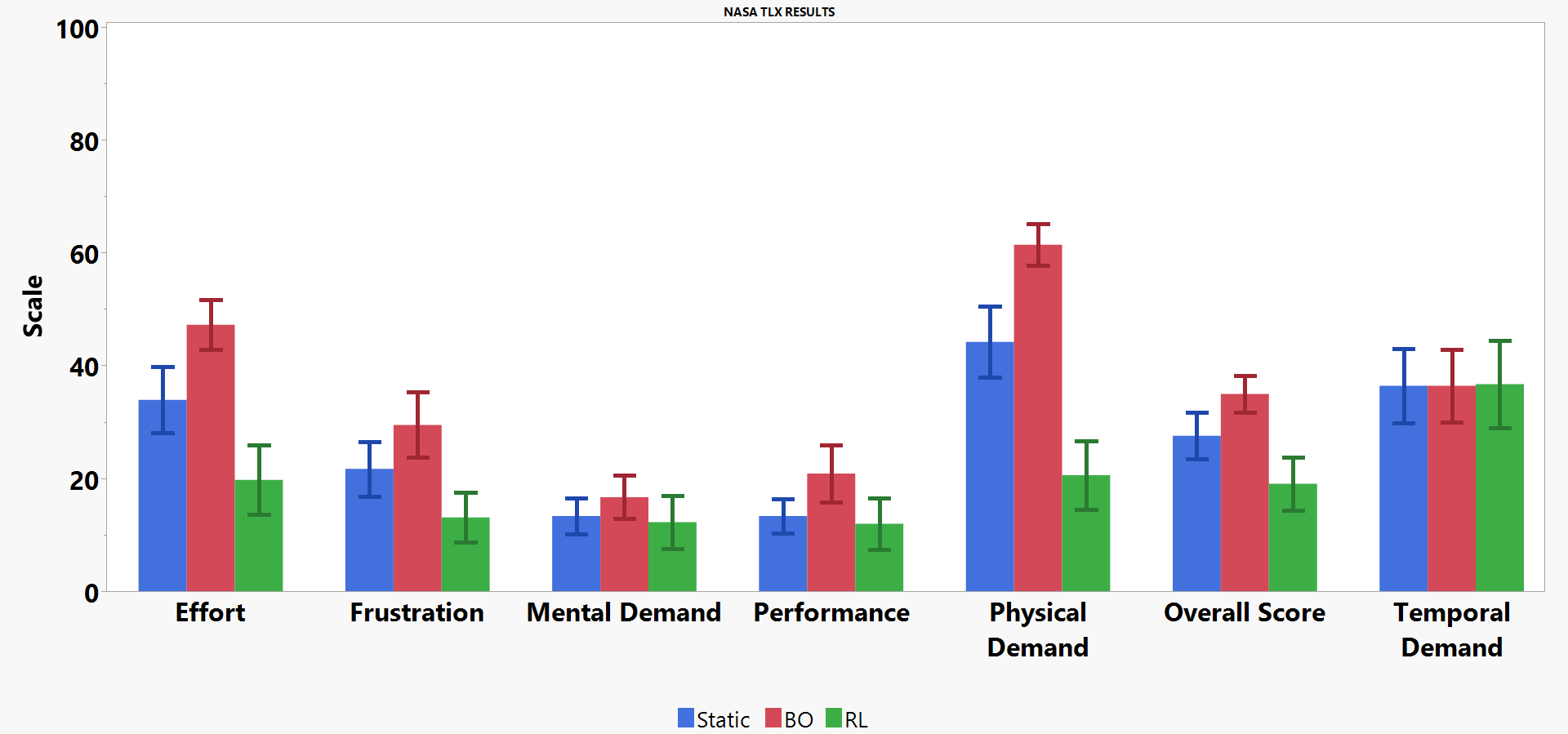}
    \caption{\textbf{NASA-TLX workload ratings across the three UI configurations.}
    Bars show mean scores for each NASA-TLX subscale, aggregated across participants.
    Lower values indicate lower perceived workload.}
    \Description{
    Bar chart showing mean NASA-TLX subscale scores for three UI configurations:  Static, Bayesian-optimized, and RL-based.
    Subscales include Mental Demand, Physical Demand, Temporal Demand, Performance, Effort, and Frustration.
    }
    \label{fig:nasa_tlx}
\end{figure}

We evaluated user workload across the three UI configurations using the NASA-TLX questionnaire (Fig.~\ref{fig:nasa_tlx}). A Friedman test on the overall NASA-TLX scores revealed a significant effect of UI configuration ($\chi^2(2) = 15.672$, $p < 0.001$).

Follow-up analyses of individual NASA-TLX subscales showed significant effects of UI configuration for \textit{Effort} ($\chi^2(2) = 12.94$, $p < 0.01$), \textit{Frustration} ($\chi^2(2) = 7.73$, $p < 0.05$), \textit{Performance} ($\chi^2(2) = 9.18$, $p = 0.01$), and \textit{Physical Demand} ($\chi^2(2) = 20.55$, $p < 0.001$).

Post-hoc pairwise comparisons were conducted using Wilcoxon signed-rank tests to identify differences between UI configurations. As shown in Table~\ref{tab:wilcoxon_nasa_subscales}, significant pairwise differences were most consistently observed between the RL-based and BO-based UIs, particularly for the Physical Demand subscale. Comparisons between the RL-based and Static UIs yielded smaller effects and were not consistently significant across subscales, with the exception of Physical Demand, where the RL-based UI resulted in significantly lower scores.

%----------------- NASA SUBSCALE TABLE --------------------------
\begin{table}[t]
\centering

\begin{tabular}{llccc}
\hline
\textbf{Subscale} & \textbf{Comparison} & \textbf{Z} & \textbf{p-value}  \\
\hline
Overall Scores &  $UI_{RL}  \text{–}  UI_{Static}$ & -2.172 &  $ < 0.05 $\\
              & $UI_{RL}  \text{–} UI_{BO} $   & -2.461 & $ < 0.05 $ \\
              & $UI_{BO} \text{–}UI_{Static} $   & -2.344 & $ < 0.05 $ \\
\hline
Effort        & $UI_{RL}  \text{–} UI_{Static}$ & -1.998 &  $ < 0.05 $ \\
              & $ UI_{RL}  \text{–} UI_{BO}  $  & -2.594 & $ < 0.01 $ \\
              & $ UI_{BO} \text{–} UI_{Static} $ & -1.827 &  n.s \\
\hline
Frustration   & $UI_{RL}  \text{–}UI_{Static} $& -1.542 &  n.s \\
              & $ UI_{RL}  \text{–} UI_{BO}$    & -2.028 & < $ 0.05 $ \\
              & $UI_{BO} \text{–} UI_{Static} $  & -1.494 &  n.s \\
\hline
Performance   & $UI_{RL}  \text{–}UI_{Static}$ & -0.910 &  n.s \\
              & $UI_{RL}  \text{–} UI_{BO} $   & -1.724 &  n.s \\
              & $UI_{BO} \text{–} UI_{Static}$   & -2.405 & $ < 0.05 $ \\
\hline
Physical Demand & $UI_{RL}  \text{–}UI_{Static}$ & -3.007 & $ < 0.01 $ \\
                 & $UI_{RL}  \text{–}UI_{BO} $   & -3.36 &  $ < 0.001 $  \\
                 &$ UI_{BO}  \text{–} UI_{Static} $& -2.421 & $ < 0.05 $ \\
\hline
\end{tabular}
\caption{Post-hoc Wilcoxon signed-rank test results for NASA-TLX subscales.}
\label{tab:wilcoxon_nasa_subscales}
\end{table}

%--------------------------------------------------------------

\section{Discussion}
\label{sec:discussion}

This work investigates whether biomechanical simulation can serve as a reliable driver for fatigue-aware VR UI layout optimization and why RL is particularly well-suited for this problem. While biomechanical models should not be interpreted as predictors of absolute human fatigue, they provide consistent and actionable signals for \emph{comparative} interface evaluation. The results also highlight the limitations of layout selection based on sparse, per-layout evaluations under noisy ergonomic feedback, motivating learning-based approaches that optimize expected cumulative interaction cost rather than isolated interface elements.

\subsection{Role of Biomechanical Models in UI Design}

A key question is whether biomechanical user models can be meaningfully integrated into UI design. The results indicate that these models are well-suited for \emph{relative} evaluation of interface layouts, even without detailed kinematic measurements or human-in-the-loop fatigue data. Simulated fatigue values may not correspond directly to perceived exertion for individual users, but the ordinal ranking of interface layouts was preserved consistently in the user study. This demonstrates that biomechanical simulation can function as an effective early-stage design tool, allowing designers to identify ergonomically unfavorable layouts and explore alternatives before conducting costly and time-consuming user studies. The value of the approach lies in guiding attention toward layouts that are likely to impose lower physical demand, rather than in providing precise fatigue predictions.

\subsection{Limitations of Bayesian Optimization under Noisy Fatigue Signals}

Despite explicitly optimizing predicted fatigue, the BO baseline underperformed both the RL-based layout and the heuristic central placement. This outcome can be attributed to the stochastic characteristics of the fatigue signal: even under fixed interaction sequences, predicted fatigue exhibits variability due to movement stochasticity and recovery dynamics in the biomechanical model. While individual measurements are noisy, the relative ordering of layouts remains stable across repeated evaluations. BO relies on a limited number of noisy evaluations to infer improvements, making it difficult to distinguish weak but systematic ergonomic advantages. 

In contrast, the RL agent optimized expected cumulative fatigue through repeated rollouts, effectively learning from distributions of outcomes rather than individual evaluations. This enabled the agent to exploit consistent ergonomic patterns such as reduced cumulative reach and compact spatial arrangements, even under stochastic per-episode feedback. The static layout benefited from a robust central-placement heuristic, while the RL agent refined beyond this baseline by explicitly optimizing cumulative interaction cost.

\subsection{Implications for VR UI Design}

The learned layouts provide concrete insights for fatigue-aware VR interface design. The RL agent consistently positioned all three buttons in the lower region of the canvas (cells 15--17), resulting in lower predicted and perceived fatigue compared to the centrally clustered static baseline (cells 8--10) and the more dispersed BO layout (cells 3, 9, and 17). From a biomechanical perspective, these placements reduce sustained shoulder elevation and upper-arm abduction across successive interactions, lowering activation of fatigue-prone muscle groups. This demonstrates that cumulative arm effort can be meaningfully reduced by leveraging row-level placement preferences rather than relying solely on small local adjustments around a central position. Importantly, these fatigue reductions were achieved without increasing task completion time, suggesting that ergonomic improvements can coexist with interaction efficiency.

The framework is designed to evaluate, compare, and refine layouts under task and sequence-specific constraints, rather than to discover a single optimal layout. Observed fatigue reductions emerge from cumulative interaction effects rather than isolated reach distances, making them difficult to anticipate through manual design alone. Training the motion and UI agents incurs upfront computation, but this cost is amortized across multiple design iterations, enabling rapid exploration of layout alternatives without repeated user studies, which are logistically more expensive and limited in coverage.

Finally, the optimization can incorporate designer-imposed constraints such as restricted regions, symmetry, accessibility requirements, or aesthetic guidelines, allowing refinement within realistic design boundaries. This demonstrates that biomechanical simulation can support iterative, informed UI design while respecting practical constraints.

\section{Application: Extending to Longer Sequential Tasks}
\label{sec:application}

The results in Section~\ref{sec:evaluation} suggest that biomechanical simulation provides a reliable \emph{comparative} signal for evaluating interface layouts, and that RL is particularly effective when ergonomic differences emerge through cumulative interaction effects rather than isolated actions. To investigate how our framework scales to more complex interactions, the three-button setup is extended to longer sequential tasks involving multiple interface elements. This setup enables evaluation of layouts under more realistic interaction patterns, where repeated selections and variable button usage affect cumulative fatigue.

\subsection{Task Definition}
The interface consists of five buttons displayed simultaneously on the UI canvas. For a given layout, the simulated user executes three button-selection sequences sampled from a predefined sequence generator, resulting in a total of nine button presses per episode. The layout remains fixed throughout each episode to reflect realistic interface behavior, as dynamically repositioning elements between sequences would be disruptive to users.

\subsubsection{Optimization Objective}
At a higher level, optimizing button placement for high-frequency usage might initially seem like a simple assignment problem. However, fatigue is cumulative and path-dependent, making it more complex than a static spatial optimization. By framing this as an RL problem, the UI agent is enabled to adapt to frequency-weighted fatigue patterns over time. In sequential interaction tasks, repeated reaching motions lead to cumulative fatigue, and elements that are used more frequently contribute significantly more to the total physical effort than less frequently used ones. The goal is to optimize interface layouts not only for spatial efficiency but also for frequency-weighted fatigue. To achieve this, the empirical button usage is tracked during each episode and incorporated into the UI agent’s reward structure. This ensures that high-frequency buttons are placed in regions that minimize physical effort, extending the optimization beyond simple spatial considerations, as seen in the three-button setup.

\subsection{RL Approach}
\subsubsection{Motion Agent}

To train the motion agent for multiple buttons simultaneously, we propose the following approach. At each step of the sequence, a small target sphere is placed at the center of the currently active button. The agent is trained to reach and select this target. After each successful interaction, the target advances to the next button in the sequence, enabling multi-step trajectories while preserving the original reward structure and motion dynamics. Following the same reward structure as in Section~\ref{sec:motionagent_rewardstruct}, the reward includes a task component $r_{\text{task}}$, a distance component $r_{\text{d}}$, and an effort regularization term \(\delta t\). A reward of 5 is given for pressing a button, and an additional reward of 15 is provided for completing the sequence. The distance reward $r_{\text{d}}$ is computed as $e^{-\mathrm{d}} - 1$, where distance is measured between the controller and the center sphere. The effort regularization term remains unchanged.

\paragraph{Sequential Interaction Modeling in VR}

\begin{figure}[t]
    \centering
    \includegraphics[height=0.275\textheight]{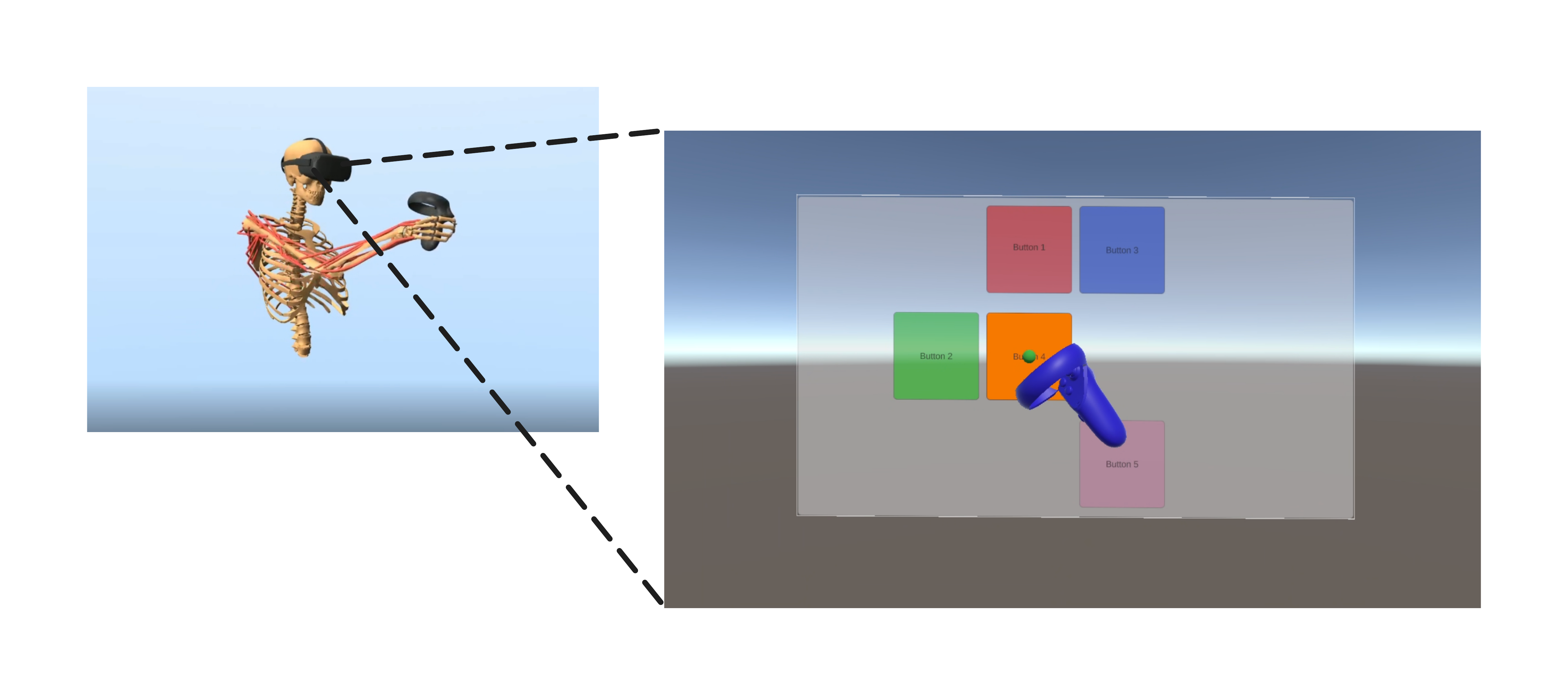}
\caption{\textbf{Sequential five-button interaction task.} All five buttons are displayed simultaneously, and the agent performs selections in the order specified by the visual marker.}

    \Description{
    Image showing the motion agent interacting with the 5-button task.
    A small spherical marker indicates the current target button.
    The simulated controller moves toward the marked button and presses it, after which the marker advances to the next button in the predefined sequence.
    The image also illustrates the low-resolution rendering (120×80 pixels) that represents the input observed by the motion agent.}
    \label{fig:five_button_interaction}
\end{figure}

In the original single-button task, reward shaping based on Euclidean distance was sufficient because only one target was visible at a time. When extending the task to multiple simultaneously visible buttons, distance-based rewards no longer reliably indicate which button in the sequence should be selected next, leading to ambiguous credit assignment. This issue is amplified by image-based observations, where multiple potential targets appear in the visual input without an explicit indication of priority. Motivated by \cite{hetzel2021complex}, we introduce an explicit target representation as a small sphere placed at the center of the currently active button. Unlike \cite{hetzel2021complex}, where the agent receives a low-dimensional state vector explicitly encoding target location and end-effector position, our motion agent operates on image-based observations and outputs muscle-level control signals, making target identity non-trivial to infer from the input alone. Alternative strategies, including directly embedding button positions or visually differentiating buttons, did not reliably disambiguate the active target. The explicit spatial target, therefore, provides a clear, visually grounded goal for sequential interactions while preserving the original reward structure (Fig.~\ref{fig:five_button_interaction}).

% To the best of our knowledge, this is among the first demonstrations of integrating a biomechanical motion model with sequential multi-target selection in a VR-based UI optimization framework.

\subsubsection{UI Agent}

The episode is defined as follows: At each timestep, the UI agent takes an action. For the same action (i.e., the same layout), three button sequences are executed, leading to nine button interactions. This design ensures that the layout does not change every timestep, avoiding confusion caused by dynamic layout changes.  

Over an episode, we record the empirical usage frequency of each button and the average fatigue cost incurred per button press. The observation space is defined as the concatenation of button usage frequencies and their current grid positions. The action space corresponds to assigning each of the five buttons to a unique cell in the discretized UI grid.

Let $n_i^{(t-1)}$ denote the number of times button $i$ was selected in the previous episode, and define the empirical usage frequency as
\[
\pi_i^{(t-1)} = \frac{n_i^{(t-1)}}{\sum_{j=1}^{5} n_j^{(t-1)}}.
\]

Empirical usage frequencies from the previous episode $\pi_i^{(t-1)}$ are used rather than those from the current rollout. Because usage statistics are only available after the episode completes, computing the reward with $\pi_i^{(t)}$ would allow the layout to influence the frequency weights used to evaluate itself, introducing circularity into the reward signal. Using $\pi_i^{(t-1)}$ instead provides a fixed estimate of expected interaction frequencies, ensuring that layout evaluation remains independent of the rollout used to measure fatigue. This design stabilizes training and avoids biased credit assignment.

Let $F_i^{(t)}$ denote the cumulative effort incurred for button $i$ during the current episode, computed as the sum of instantaneous effort costs $C_{\mathrm{eff}}(t)$ over all transitions to that button.

The reward at episode $t$ is then defined as
\[
R_t = e^{-\sum_{i=1}^{5} \pi_i^{(t-1)} \, F_i^{(t)}}.
\]

An exponential scaling is applied to increase the sensitivity of the reward to fatigue incurred by frequently used buttons. This ensures that layouts placing high-frequency buttons in moderately high-cost regions are penalized more strongly, providing a clearer optimization signal for the UI agent. In contrast to the previous three-button setup, where linear scaling of cost was sufficient, the current design involves nine button interactions per episode. Using a linear combination of frequency and cost would produce relatively small differences in reward values, which could hinder PPO's learning by generating weak gradient signals. The exponential formulation amplifies these differences, allowing the agent to more effectively distinguish between layouts and accelerate convergence during training.

To avoid overlapping buttons, a large penalty is applied whenever multiple buttons occupy the same grid cell. This ensures that invalid UI placements are clearly discouraged.

\begin{figure}[t]
    \centering
    \includegraphics[width=0.5\linewidth]{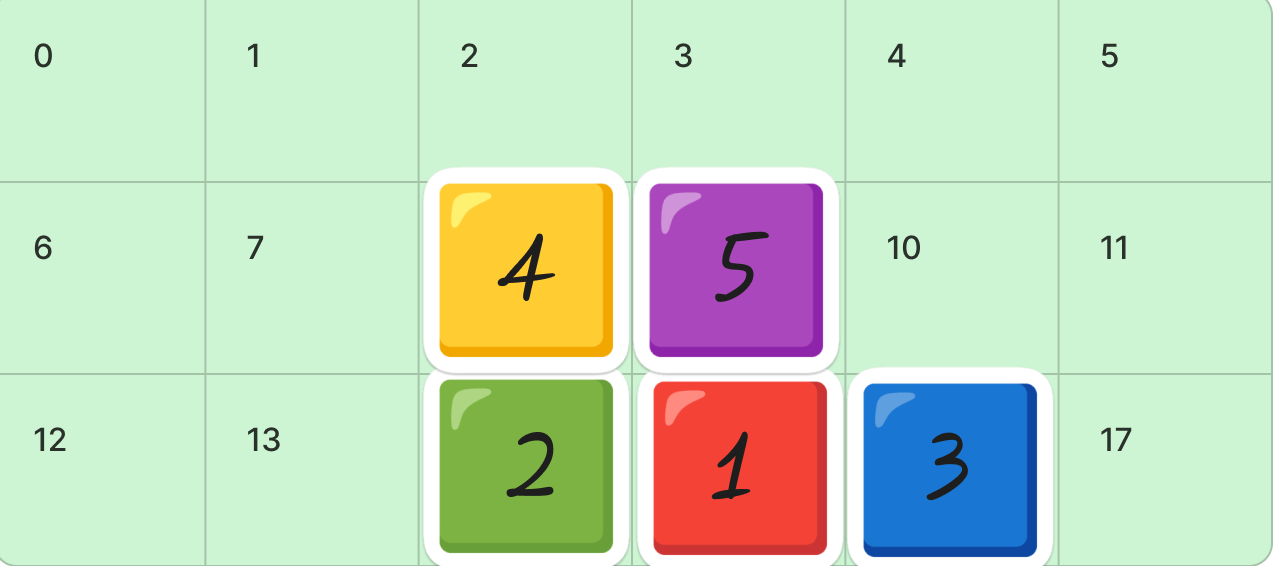}
    \caption{\textbf{Five-button layout optimized under frequency-weighted fatigue.}
    Frequently used buttons are placed in lower-cost regions of the canvas, reflecting the agent’s adaptation to cumulative biomechanical effort across sequential interactions.}
    \Description{
    Visualization of a five-button UI layout on a discretized grid.
    Five buttons occupy distinct grid cells on the canvas, illustrating a layout selected by the RL-based UI agent.
    The layout is intended to place more frequently used buttons in lower-cost regions under the simulated fatigue model.
    }
    \label{fig:ui_rl_five_buttons}
\end{figure}

\subsubsection{Results}

Across episodes, the UI agent consistently converged to a single layout configuration, even when button usage frequencies varied (Fig.~\ref{fig:ui_rl_five_buttons}). This behavior is a consequence of the optimization formulation rather than a limitation of frequency-aware learning. In the reward formulation, each button’s fatigue cost is scaled by how often it is used, so frequently pressed buttons contribute more to the overall penalty. However, the relative difficulty of positions on the canvas remains largely stable, meaning that frequency weighting amplifies penalties but does not change which locations are easiest or hardest to reach. As a result, the agent learns a layout that minimizes expected fatigue across all episodes, rather than adjusting button placements to per-episode usage fluctuations. This illustrates that policy-based optimization finds a globally efficient layout under the expected interaction statistics, highlighting the value of incorporating frequency into the fatigue signal. This demonstrates that the framework proposed produces ergonomically sensible layouts at scale and validates the utility of frequency-weighted fatigue as a design signal, even when per-episode variations are small.

\section{Limitations and Future Work}
\label{sec:limitations_futurework}

This extension to longer sequential tasks illustrates the role of cumulative fatigue as an effective optimization signal and shows that the proposed framework remains well-behaved under more dynamic and realistic interaction scenarios. The focus of this analysis is on scalability and optimization behavior as task complexity increases.

The results should therefore be interpreted at the level of the simulated model. While the predicted fatigue signals were effective for comparative evaluation of interface layouts, they are not intended to serve as absolute predictors of perceived exertion for specific users. Biomechanical models provide a reasonable baseline by mimicking human movement strategies, but they lack cognitive processes and decision-making, and cannot fully capture adaptive or nuanced human behaviors. In addition, individual differences in anthropometry, posture, strength, and movement strategy were not explicitly modeled. This abstraction enables controlled analysis of physical effort, but real-world VR fatigue may also be influenced by cognitive load, visual strain, or prior physical activity—factors not captured by the current simulation.

Future work could address these limitations by introducing variability in anthropometric parameters, such as sampling different arm lengths during execution, to better approximate user diversity. In this work, we focused on a right-handed upper-body model, which limits generalizability to left-handed or ambidextrous users. Extending the framework to full-body or bimanual biomechanical models would allow investigation of fatigue patterns in larger or more complex interfaces involving both controllers \cite{caggiano2022myosuite, wang2022myosim}. Such extensions would also support the study of multi-step or multi-task interaction scenarios in which fatigue accumulates across broader task contexts.

From a systems perspective, our architecture follows a hierarchical RL structure in which the UI agent depends on a pretrained biomechanical motion agent. Although the motion agent generalizes to minor variations in the VR task (e.g., button size or position), its ability to adapt to substantial changes in interaction mechanics or interface design is limited. Future work could explore curriculum learning, domain randomization, or fine-tuning strategies to improve generalization across interface variations without requiring full retraining from scratch.

Finally, while we demonstrated the framework using PPO, the proposed formulation is not tied to a specific learning algorithm. Alternative optimization methods or hybrid approaches could be explored depending on design constraints and interaction context. The framework could also be extended to support more complex or adaptive UI elements, such as dynamically adjusting layouts, variable button sizes, or compound widgets (e.g., drop-down menus), enabling investigation of scalability in more flexible and user-responsive interface designs.

\section{Conclusion}
\label{sec:conclusion}

We present a framework that systematically integrates biomechanical simulation into the VR UI design process to support fatigue-aware layout optimization. Although simulated fatigue should not be treated as an absolute predictor of human effort, our results demonstrate that biomechanical models provide consistent \emph{comparative} signals for evaluating and refining interface layouts. By framing layout selection as a sequence-level optimization problem under noisy ergonomic feedback, we show that RL is well-suited to optimizing layouts under cumulative fatigue signals that are difficult to address with static or single-shot approaches. Rather than identifying a single universally optimal layout, the framework enables early-stage exploration of ergonomically informed design alternatives, reducing reliance on costly user studies and facilitating the development of more physically considerate VR interfaces.

\section*{Acknowledgement}
This research was undertaken, in part, based on support from the Natural Sciences and Engineering Research Council of Canada Grant RGPIN-2021-03479 (NSERC DG).
% \clearpage
\bibliographystyle{ACM-Reference-Format}
\bibliography{sample-base}

\end{document}